\documentclass[showpacs,10pt,twocolumn,prb]{revtex4-1}

\usepackage{amsmath}
\usepackage{amssymb}
\usepackage{graphicx}
\usepackage{amssymb}
\usepackage{graphics}
\usepackage{epsfig}
\usepackage{CJK}
\usepackage{color}

\begin{document}

\begin{CJK*}{GBK}{Song}
\title{Anomalous Hall effect in van der Waals bonded ferromagnet Fe$_{3-x}$GeTe$_2$}
\author{Yu Liu,$^{1}$ Eli Stavitski,$^{2}$ Klaus Attenkofer,$^{2}$ and C. Petrovic$^{1}$}
\affiliation{$^{1}$Condensed Matter Physics and Materials Science Department, Brookhaven National Laboratory, Upton, New York 11973, USA\\
$^{2}$National Synchrotron Light Source II, Brookhaven National Laboratory, Upton, New York 11973, USA
}
\date{\today}

\begin{abstract}
We report anomalous Hall effect (AHE) in single crystals of quasi-two-dimensional Fe$_{3-x}$GeTe$_2$ ($x \approx 0.36$) ferromagnet grown by the flux method which induces defects on Fe site and bad metallic resistivity. Fe K-edge x-ray absorption spectroscopy was measured to provide information on local atomic environment in such crystals. The dc and ac magnetic susceptibility measurements indicate a second-stage transition below 119 K in addition to the paramagnetic to ferromagnetic transition at 153 K. A linear scaling behavior between the modified anomalous Hall resistivity $\rho_{xy}/\mu_0H_{eff}$ and longitudinal resistivity $\rho_{xx}^2M/\mu_0H_{eff}$ implies that the AHE in Fe$_{3-x}$GeTe$_2$ should be dominated by the intrinsic Karplus-Luttinger mechanism rather than the extrinsic skew-scattering and side-jump mechanisms. The observed deviation in the linear-M Hall conductivity $\sigma_{xy}^A$ below 30 K is in line with its transport characteristic at low temperatures, implying the scattering of conduction electrons due to magnetic disorder and the evolution of the Fermi surface induced by possible spin-reorientation transition.
\end{abstract}

\maketitle
\end{CJK*}

\section{INTRODUCTION}

In addition to the ordinary Hall effect originating from the deflection of moving charge carriers by the Lorentz force in magnetic field, anomalous Hall effect (AHE) proportional to the spontaneous magnetization $M$ arises in magnetic materials. This is of considerable interest in fundamental physics and applied sciences alike.\cite{Webster, Bergmann, Nagaosa, Wang, Manyala, Husmann} Three mechanisms responsible for the AHE are widely accepted. The intrinsic Karplus and Luttinger (KL) mechanism is related to the spin-orbit coupling (SOC) and perturbation by the applied electric field, resulting in an additional term in the carrier group velocity.\cite{Nagaosa, Karplus} The extrinsic mechanisms involving the skew-scattering and side-jump mechanism can also give rise to the AHE and are induced by asymmetric scattering of conduction electrons.\cite{Smit, Berger}

Two-dimensional (2D) materials such as graphene and transition-metal dichalcogenides exhibit a number of attractive properties that have been extensively studied in the past two decades.\cite{Geim, Hu, Bhimanapati} In contrast to the mechanical and optoelectronic properties, however, magnetism in 2D materials has received little attention until recently.\cite{MillerJ,HuangB,SamarthN} Van der Waals (VDW) bonded magnetic materials are of great interest as building blocks for heterostructures in spin-based information technologies. Chromium-based CrX$_3$ (X = Cl, Br, I) and CrXTe$_3$ (X = Si, Ge, Sn) have been identified as the promising candidates for long-range magnetism in nanosheets.\cite{Zhang1, McGuire, Sivadas, Zhuang1, Lin1} CrSiTe$_3$ exhibits ferromagnetic (FM) order below 32 K in bulk,\cite{Casto} and $\sim$ 80 K in monolayer and few-layer samples.\cite{Lin2} Bulk CrI$_3$ and CrGeTe$_3$ exhibit FM below 61 K.\cite{McGuire, Zhang2} Fe$_{3-x}$GeTe$_2$ is of particular interest due to higher Curie temperature ($T_c$) and possibility for tuning of magnetism by Fe defects and site occupancies control by different synthesis routes.\cite{Deiseroth,Andrew}

The ternary Fe$_{3-x}$GeTe$_2$ is a layered 2D material with Fe$_{3-x}$Ge slabs sandwiched between two VDW bonded Te layers, which was first synthesized by Abribosov \emph{et al.}.\cite{Abribosov} Fe$_{3-x}$GeTe$_2$ is a weak itinerant ferromagnet with the $T_c$ of 220 K and competing antiferromagnetic (AFM) interaction along the c axis below 152 K reported for $x$ = 0; however ferromagnetic $T_c$ decreases with an increase in Fe vacancies.\cite{Deiseroth, Zhu, Chen1, Yi, Andrew, Abribosov} The Fe atoms in the unit cell occupy two inequivalent Wyckoff sites, i. e., the Fe1 atoms are situated in hexagonal net layer with only Fe atoms, whereas Fe2 and Ge atoms are covalently bonded in an adjacent layer. The flux-grown crystals typically have a lower $T_c$ of 150 K with Fe vacancies level $x \approx 0.3$,\cite{Andrew} but in such crystals Fe vacancies are only present in the Fe2 atomic sites whereas no Fe atoms occupy in interlayer space.\cite{Andrew, Yu} Furthermore, the density-functional calculations predict that the single-layer Fe$_3$GeTe$_2$ is dynamically stable, and that it exhibits a significant uniaxial magnetocrystalline anistropy energy, potentially useful for magnetic storage applications.\cite{Zhuang2}

Here we present a study of AHE in the flux-grown single crystals of Fe$_{3-x}$GeTe$_2$, in connection with its magnetic and transport properties. Zero-field-cooling (ZFC) and field-cooling (FC) curves of dc magnetization exhibit significant splitting at low temperatures for $H//c$, but not for $H//ab$, in line with its large magnetic anisotropy. A second-stage transition below 119 K in addition to $T_c = 153$ K is confirmed by ac susceptibility. The linear dependence of the modified anomalous Hall resistivity $\rho_{xy}/\mu_0H_{eff}$ and longitudinal resistivity $\rho_{xx}^2M/\mu_0H_{eff}$ indicates that the intrinsic KL mechanism dominates the AHE in Fe$_{3-x}$GeTe$_2$.

\section{EXPERIMENTAL DETAILS}

Single crystals of Fe$_{3-x}$GeTe$_2$ were grown by the flux method.\cite{YULIU} The stoichiometry was measured by examination of multiple points using x-ray energy-dispersive spectroscopy (EDS) with a JEOL LSM-6500 scanning electron microscope (SEM). The x-ray absorption spectroscopy (XAS) measurements were performed at 8-ID beamline of the National Synchrotron Light Source II (NSLS II) at Brookhaven National Laboratory (BNL) in the transmission mode. The x-ray absorption near edge structure (XANES) and extended x-ray absorption fine structure (EXAFS) spectra were processed using the Athena software package. The AUTOBK code was used to normalize the absorption coefficient, and separate the EXAFS signal, $\chi(k)$, from the atom-absorption background. The extracted EXAFS signal, $\chi(k)$, was weighed by $k^2$ to emphasize the high-energy oscillation and then Fourier-transformed in a $k$ range from 2 to 12 {\AA}$^{-1}$ to analyze the data in $R$ space. The dc/ac magnetic susceptibility, electrical and thermal transport, and heat capacity were measured in the Quantum Design MPMS-XL5 and PPMS-9 systems. The longitudinal and Hall resistivity were performed using a standard four-probe method with the current flowing in the $ab$ plane of hexagonal structure. In order to effectively eliminate the longitudinal resistivity contribution due to voltage probe misalignment, the Hall resistivity was obtained by the difference of transverse resistance measured at positive and negative fields, i.e., $\rho_{xy}(\mu_0H) = [\rho(+\mu_0H)-\rho(-\mu_0H)]/2$.

\section{RESULTS AND DISCUSSIONS}

\begin{figure}
\centerline{\includegraphics[scale=0.8]{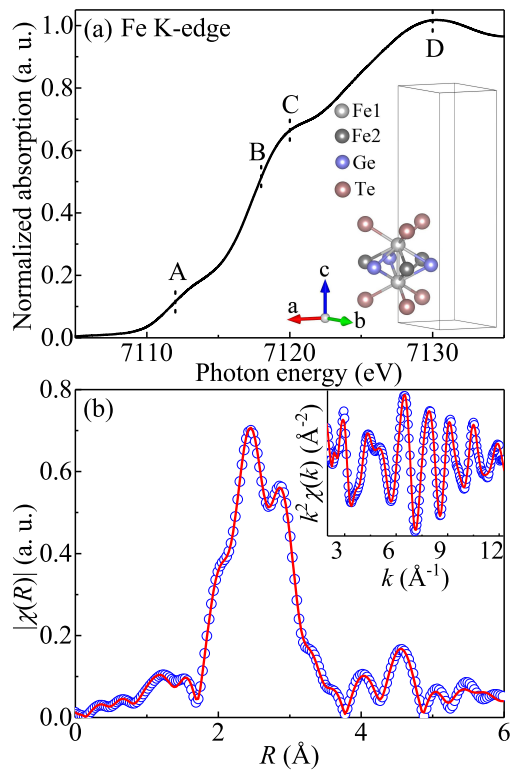}}
\caption{(Color online). Normalized Fe K-edge XANES spectra (a) and Fourier transform magnitudes of EXAFS data (b) of Fe$_{3-x}$GeTe$_2$ measured at room temperature. The experimental data are shown as blue symbols alongside the model fit plotted as red line. The inset in (b) shows the corresponding EXAFS oscillation with the model fit.}
\label{EXAFS}
\end{figure}

\begin{table}
\caption{\label{tab1}Local structural parameters extracted from the Fe K-edge EXAFS spectra of Fe$_{3-x}$GeTe$_2$. CN is coordination number based on crystallographic value, R is interatomic distances, and $\sigma^2$ is Debye Waller factor. }
\begin{ruledtabular}
\begin{tabular}{llll}
   & CN & R ({\AA}) & $\sigma^2$ ({\AA}$^2$)\\
  \hline
  Fe1-Fe1 & 1 & 2.55(39) & 0.001(12) \\
  Fe1-Fe2 & 3 & 2.61(4) & 0.01(1) \\
  Fe1-Ge & 3 & 2.61(4) & 0.02(6)\\
  Fe1-Te & 3 & 2.64(6) & 0.07(6)\\
  Fe1-Fe1 & 6 & 3.94(49) & 0.04(2) \\
  Fe1-Te & 3 & 4.51(8) & 0.018(6) \\
  Fe1-Fe1 & 6 & 4.70(2) & 0.04(1)\\
  Fe1-Ge & 3 & 4.73(22) & 0.02(1)\\
  Fe1-Fe2 & 3 & 4.73(22) & 0.02(1)
\end{tabular}
\end{ruledtabular}
\end{table}

The EDS gives a composition of Fe$_{2.64(6)}$Ge$_{0.87(4)}$Te$_2$ in our flux-grown single crystals with Fe deficiency $x \approx 0.36$, in good agreement with the previous report.\cite{Andrew} The average crystal structure and x-ray diffraction (XRD) of single and crushed crystals were reported in our previous paper.\cite{YULIU} Figure \ref{EXAFS} shows the normalized Fe K-edge XANES spectra and Fourier transform magnitudes of EXAFS spectra of Fe$_{3-x}$GeTe$_2$ obtained at room temperature. The typical features of near edge are marked as A, B, C, and D [Fig. \ref{EXAFS}(a)]. The prepeak feature A ($\sim$ 7112 eV) is the result of a direct quadrupole transition to unoccupied 3d states that are hybridized with Te 4p orbitals. The edge feature B ($\sim$ 7118 eV) is governed by 1s $\rightarrow$ 4p transition, which is close to the one for a reference Fe$^{2+}$ standard,\cite{Chang, Joseph} indicating the Fe$^{2+}$ state. The peak-like feature C ($\sim$ 7120 eV) should be driven by the 1s $\rightarrow$ 4p states admixed with Te d states, and the feature D is mainly due to multiple scattering of the photoelectrons with the nearest neighbors. In the single-scattering approximation, the EXAFS could be described by the following equation\cite{Prins}
\begin{align*}
\chi(k) = \sum_i\frac{N_iS_0^2}{kR_i^2}f_i(k,R_i)e^{-\frac{2R_i}{\lambda}}e^{-2k^2\sigma_i^2}sin[2kR_i+\delta_i(k)],
\end{align*}
where $N_i$ is the number of neighbouring atoms at a distance $R_i$ from the photoabsorbing atom. $S_0^2$ is the passive electrons reduction factor, $f_i(k, R_i)$ is the backscattering amplitude, $\lambda$ is the photoelectron mean free path, $\delta_i$ is the phase shift of the photoelectrons, and $\sigma_i^2$ is the correlated Debye-Waller factor measuring the mean square relative displacement of the photoabsorber-backscatter pairs. In present case, the first nearest neighbors of Fe1 atoms are Fe1, 3Fe2, 3Ge, and 3Te atoms located at 2.55 {\AA} $\sim$ 2.65 {\AA} [inset in Fig. \ref{EXAFS}(a)], and the next nearest neighbors are 6Fe1, 3Te, 6Fe1, 3Ge, and 3Fe2 atoms sited at 3.95 {\AA} $\sim$ 4.75 {\AA}.\cite{Andrew} Local structural information, such as the bond distance and Debye-Waller factor, were obtained by the best-fit model involving the near neighbors of Fe1 atoms below 5 {\AA} [Fig. \ref{EXAFS}(b)]. The parameters are summarized in Table I. The stoichiometry of Fe$_{2.64(6)}$Ge$_{0.87(4)}$Te$_2$ with Fe deficiency $x \approx 0.36$ confirmed in our flux-grown single crystals, in good agreement with the previous report,\cite{Andrew} makes its footprint in the broadening of the first main peak. The features above 5 {\AA} are due to longer distances and multiple scattering effects.

\begin{figure}
\centerline{\includegraphics[scale=0.825]{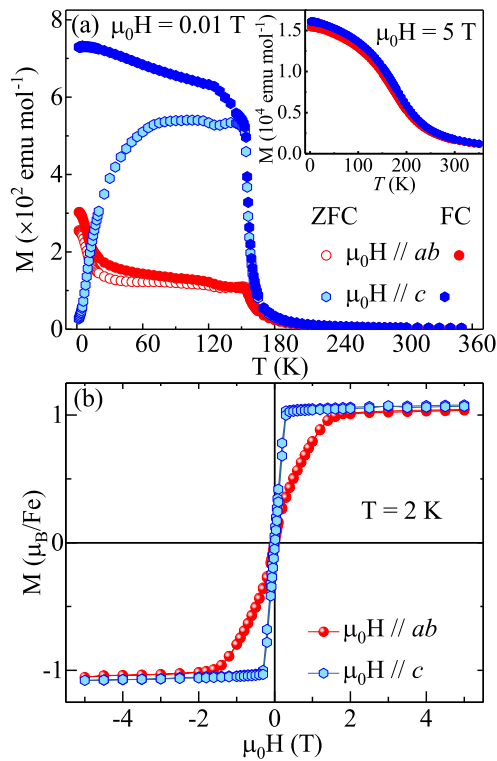}}
\caption{(Color online). (a) Temperature dependence of dc magnetization $M(T)$ with zero-field cooling (ZFC) and field-cooling (FC) taken at $\mu_0H$ = 0.01 T and $\mu_0H$ = 5 T (inset) for $\mu_0H//ab$ and $\mu_0H//c$, respectively. (b) Field dependence of magnetization $M(\mu_0H)$ taken at $T$ = 2 K.}
\label{MT}
\end{figure}

Figure \ref{MT}(a) shows the temperature dependence of dc magnetization $M(T)$ measured at low field $\mu_0H$ = 0.01 T applied in the $ab$ plane and parallel to the $c$ axis, respectively. An obvious paramagnetic (PM) to FM transition was observed, followed by an additional weak kink just below that, suggesting a two-stage magnetic ordering behavior. In addition, the ZFC and FC curves show significant splitting at low temperatures for $\mu_0H//c$, but not for $\mu_0H//ab$, in line with its large magnetic anisotropy. Below 30 K, the ZFC magnetization along c axis (easy axis) is even lower than the counterpart in the ab plane (hard axis), indicating different spin coupling and/or spin reorientation at low temperatures. The isothermal magnetization $M(\mu_0H)$ taken at $T$ = 2 K for $\mu_0H//ab$ and $\mu_0H//c$ is shown in Fig. \ref{MT}(b). The saturation field $H_s \approx 0.3 $ T for $\mu_0H//c$ is much smaller than $H_s \approx 1.8 $ T for $\mu_0H//ab$. The estimated saturation moments at $T$ = 2 K are $M_s \approx$ 1.00(1) $\mu_B/$Fe for $\mu_0H//ab$ and $M_s \approx$ 1.03(1) $\mu_B/$Fe for $\mu_0H//c$, in good agreement with previous reports.\cite{Deiseroth, Zhu, Chen1, Yi, Andrew, Abribosov}

\begin{figure}
\centerline{\includegraphics[scale=0.9]{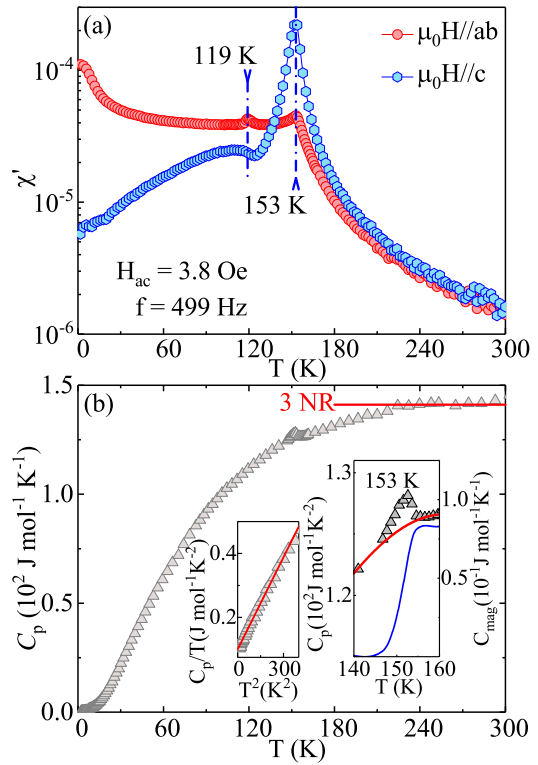}}
\caption{(Color online). (a) Temperature dependence of ac susceptibility real part $\chi^\prime(T)$ measured in zero external field. Oscillated ac field of 3.8 Oe is applied in the $ab$ plane and the $c$ axis, respectively. (b) Temperature dependence of heat capacity $C_p(T)$. Insets: The low temperature $C_p(T)/T$ vs $T^2$ curve fitted by $C_p(T)/T = \gamma + \beta T^2$ and the enlargement of the $\lambda$-type anomaly around $T_c$ = 153 K. The red curve in right inset represents the phonon contribution fitted by a polynomial. The right axis and its associated blue curve denote the magnetic entropy calculated by $S_{mag}(T) = \int_0^TC_{mag}/TdT$.}
\label{Cp}
\end{figure}

To determine the accurate transition temperatures, the ac magnetic susceptibility was measured at oscillated ac field of 3.8 Oe and frequency of 499 Hz. Two peaks in the real part $\chi^\prime(T)$ [Fig. \ref{Cp}(a)]: the PM-FM transition at 153 K and an additional weak peak at 119 K, confirm it is a two-step magnetic ordering. Neutron scattering and/or magnetic force microscopy are needed to further clarify its mechanism. In stoichiometric Fe$_3$GeTe$_2$ crystal grown by chemical vapor transport (CVT),\cite{Yi} a similar behavior was observed at higher temperatures (214 K and 152 K), in agreement with tunable magnetism by different synthesis routes. Figure \ref{Cp}(b) shows the temperature dependence of heat capacity $C_p(T)$ for Fe$_{3-x}$GeTe$_2$, in which a clear $\lambda$-type anomaly was observed at 153 K, consistent with the PM-FM transition. The high temperature $C_p(T)$ approaches the Dulong Petit value of $3NR$ $\approx$ 137 J mol$^{-1}$ K$^{-1}$, where $R$ is the molar gas constant. The low temperature data from 2 K to 16 K are featureless, suggesting the absence of Kondo scattering contribution to resistivity upturn at low temperatures (see the discussion below) and can be well fitted by $C_p(T)/T = \gamma + \beta T^2$, where the first term is the Sommerfeld electronic specific heat coefficient and the second term is low-temperature limit of the lattice heat capacity, as shown in the left inset of Fig. \ref{Cp}(b). The obtained $\gamma$ and $\beta$ are 100(1) mJ mol$^{-1}$ K$^{-2}$ and 0.962(8) mJ mol$^{-1}$ K$^{-4}$, respectively. The Debye temperature $\Theta_D$ = 230(1) K can be derived from $\beta$ using $\Theta_D = (12\pi^4NR/5\beta)^{1/3}$, where $N$ is the number of atoms per formula unit. Magnetic contribution ($C_{mag}$) can be obtained after subtraction of the phonon contribution ($C_{ph}$) fitted using a polynomial. Then the magnetic entropy can be calculated by $S_{mag}(T) = \int_0^TC_{mag}/TdT$. The derived $S_{mag}$ is $\sim$ 0.083 J mol$^{-1}$ K$^{-1}$ when $T$ is up to 160 K, which is only $\sim$ 1.4$\%$ $Rln2$ for $S = 1/2$, suggesting possible short-range order which partially releases the magnetic entropy in addition to long-range magnetic transition.

\begin{figure}
\centerline{\includegraphics[scale=0.8]{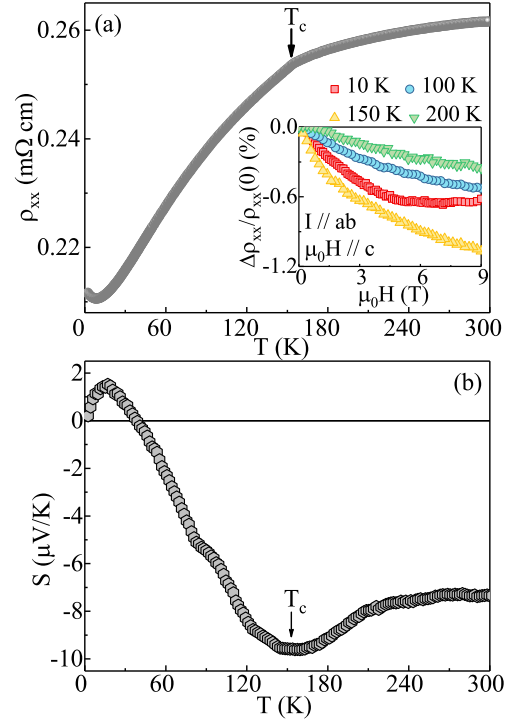}}
\caption{(Color online). (a) Temperature-dependent in-plane resistivity $\rho_{xx}(T)$ for Fe$_{3-x}$GeTe$_2$. Inset: Magnetoresistance of $\rho_{xx}(\mu_0H)$ at different temperatures. (b) Temperature-dependent Seebeck coefficient $S(T)$ for Fe$_{3-x}$GeTe$_2$.}
\label{RSMHT}
\end{figure}

Figure \ref{RSMHT}(a) shows the temperature-dependent in-plane resistivity $\rho_{xx}(T)$ of Fe$_{3-x}$GeTe$_2$, indicating bad metallic behavior with a clear kink at $T_c = $ 153 K and a weak upturn below 15 K. Magnetoresistance measured with magnetic field $\mu_0H//c$ and the current flowing in the $ab$ plane, $MR = [\rho_{xx}(\mu_0H)-\rho_{xx}(0)]/\rho_{xx}(0)$, is negative in the whole temperature range with a maximum $\sim -1.1\%$ at 150 K [inset in Fig. \ref{RSMHT}(a)], due to suppression of spin scattering in ferromagnetic Fe$_{3-x}$GeTe$_2$ by magnetic field. The temperature-dependent Seebeck coefficient $S(T)$ is negative at high temperatures with a maximum value of -9.6 $\mu$V K$^{-1}$ around 153(8) K, consistent with the anomaly in the resistivity data and indicating dominant negative charge carriers [Fig. \ref{RSMHT}(b)]. With further temperature decrease, the absolute value of $S(T)$ decreases gradually and then changes its sign to positive below 39 K with a maximum around 15 K, implying possible multi-band transport. The $S(T)$ peak around 15 K and weak increase in $\rho(T)$ below the same temperature range on cooling probably arise due to frozen-in defect-induced randomness, similar to quasicrystals.\cite{FisherIR}

\begin{figure}
\centerline{\includegraphics[scale=0.95]{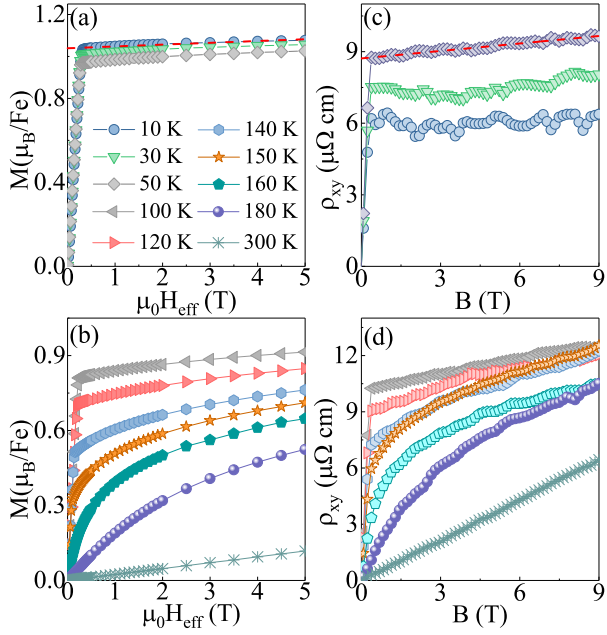}}
\caption{(Color online). (a,b) Effective field dependence of magnetization $M(\mu_0H_{eff})$ and (c,d) Hall resistivity $\rho_{xy}(B)$ for Fe$_{3-x}$GeTe$_2$ at various temperatures with $\mu_0H//c$. The red dashed lines in (a) and (c) are linear fits of $M(\mu_0H_{eff})$ and $\rho_{xy}(B)$ at high field region.}
\label{MHHall}
\end{figure}

Figures \ref{MHHall}(a,b) show the effective field dependence of magnetization at various temperatures for $\mu_0H//c$. Here $\mu_0H_{eff} = \mu_0(H-N_dM)$, where $N_d$ is the demagnetizing factor. For a sample with dimensions 2.31 mm $\times$ 3.32 mm $\times$ 0.086 mm, the calculated value of $N_d =$ 0.9.\cite{Aharoni} When $T < T_c$, the shape of $M(\mu_0H_{eff})$ curves is typical for ferromagnets, i.e., a rapid increase at low field region with a saturation in higher magnetic fields. The saturation magnetization $M_s$ decreases with increasing temperature, consistent with the temperature-dependent $M(T)$ [Fig. 2(a)] as well as the trend of magnetic moment obtained from neutron powder diffraction (NPD).\cite{Andrew} When $T > T_c$, it gradually changes into linear-in-field paramagnetic dependence. Hall resistivity $\rho_{xy}(B)$ as a function of magnetic induction $B$ for Fe$_{3-x}$GeTe$_2$ at the corresponding temperatures are depicted in Figs. \ref{MHHall}(c,d). Here $B = \mu_0(H_{eff}+M) = \mu_0[H + (1-N_d)M]$ with $N_d = 0.9$. When $T < T_c$, the $\rho_{xy}(B)$ increases quickly at low $B$ region. With increasing $B$, the $\rho_{xy}(B)$ curve changes slightly with almost linear $B$ dependence at high $B$ region, similar to the shape of $M(\mu_0H_{eff})$ curve, indicating an AHE in Fe$_{3-x}$GeTe$_2$.

\begin{figure}
\centerline{\includegraphics[scale=1]{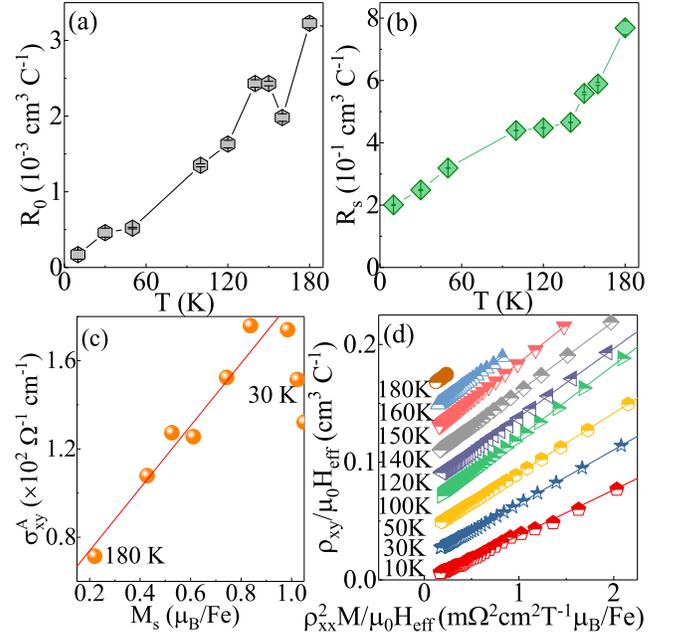}}
\caption{(Color online). (a,b) Temperature-dependent $R_0(T)$ and $R_s(T)$ fitted from $\rho_{xy}(B,T)$ using $\rho_{xy} = R_0B + R_s\mu_0M$. (c) Anomalous Hall conductivity $\sigma_{xy}^A$ vs $M_s$. (d) The $\rho_{xy}/\mu_0H_{eff}$ vs $\rho_{xx}^2M/\mu_0H_{eff}$ curves at indicated temperatures with subsequent offset of 0.02 cm$^3$ C$^{-1}$. The solid lines in (d) represent linear fits of data at different temperatures.}
\label{parameters}
\end{figure}

In general, the Hall resistivity $\rho_{xy}$ in the ferromagnets is made up of two parts,\cite{Wang, WangY, Onoda2008}
\begin{align*}
\rho_{xy} = \rho_{xy}^O + \rho_{xy}^A = R_0B + R_s\mu_0M,
\end{align*}
where $\rho_{xy}^O$ and $\rho_{xy}^A$ are the ordinary and anomalous Hall resistivity, and $R_0$ and $R_s$ are the ordinary and anomalous Hall coefficient, respectively. A linear fit of $\rho_{xy}(B)$ at high field region, the slope and $y$ axis intercept corresponds to $R_0$ and $\rho_{xy}^A$, respectively. As shown in Fig. \ref{parameters}(a), the values of $R_0$ are positive, in contrast with the negative value of thermoelectric power $S(T)$, indicating multiple carriers transport. The sign of $S(T)$ changes since there is different dependence on carrier density $n_e$ ($n_h$), mobility $\mu_e$ ($\mu_h$), and $S_e$ ($S_h$) in two-band model [$S = (S_en_e\mu_e + S_hn_h\mu_h)/(n_e\mu_e + n_h\mu_h)$]. On the other hand, the value of $R_s$ can be obtained by using $\rho_{xy}^A = R_s \mu_0 M_s$ with $M_s$ taken from the linear fit of $M(\mu_0H_{eff})$ curves at high field region, which decreases monotonically with decreasing temperature. Additionally, the value of $R_s$ is two orders of magnitude larger than that of $R_0$. Given a weak temperature-dependent resistivity of 0.24 $\pm$ 0.02 m$\Omega$ cm [Fig. \ref{RSMHT}(a)], the estimated carrier concentration $n\sim10^{22}$ cm$^{-3}$ points to a mean free path $\lambda \sim$ 0.10(1) nm, comparable to the lattice parameters and close to the Mott-Ioffe-Regel limit.\cite{GunnarsonO} This is in agreement with its bad metal behavior.

Three possible mechanisms are considered to explain the AHE. The KL mechanism has lately been reinterpreted through a Berry curvature term, which is an intrinsic property of the occupied electronic states in a crystal with certain symmetry.\cite{Jungwirth, Onoda}  It takes nonzero value only in systems where time-reversal symmetry is broken or where net magnetic moments are present, producing the scaling behavior of $\rho_{xy}^A = \beta\rho_{xx}^2$. The side-jump mechanism, where the potential field induced by impurities contributes to the anomalous group velocity, follows the same quadratic scaling behavior with the KL mechanism. However, the skew-scattering mechanism which describes asymmetric scattering induced by impurity or defect could contribute to the AHE with scaling behavior of $\rho_{xy}^A = \beta\rho_{xx}$.

The anomalous Hall conductivity $\sigma_{xy}^A$ ($\approx$ $\rho_{xy}^A / \rho_{xx}^2$) is shown in Fig. 6(c). Theoretically, the intrinsic contribution of $\sigma_{xy,in}^A$ is of the order of $e^2/(ha)$, where $e$ is the electronic charge, $h$ is the Plank constant, and $a$ is the lattice parameter.\cite{Onoda2006} Taking $a = V^{1/3} \sim 6.0$ {\AA} approximately, the $\sigma_{xy,in}^A$ is $\sim$ 646 $\Omega^{-1} cm^{-1}$. It is very close to the value of stoichiometric Fe$_3$GeTe$_2$,\cite{WangY} larger than but still the same magnitude of order of the calculated $\sigma_{xy}^A$ of Fe$_{3-x}$GeTe$_2$ [Fig. 6(c)]. By contrast, the extrinsic side-jump contribution of $\sigma_{xy,sj}^A$ has been shown to be on the order of $e^2/(ha)(\varepsilon_{SO}E_F)$, where $\varepsilon_{SO}$ and $E_F$ is the SOI and Fermi energy, respectively.\cite{Nozieres} Since the $\varepsilon_{SO}E_F$ is usually less than $10^{-2}$ for the metallic ferromagnets, the extrinsic side-jump contribution should be small and the AHE of Fe$_{3-x}$GeTe$_2$ is dominated by the intrinsic KL contribution. For the intrinsic AHE, the $\sigma_{xy}^A$ is proportional to $M$,\cite{Manyala} the scaling coefficient $S_H = \mu_0R_s/\rho_{xx}^2 = \sigma_{xy}^A/M_s$ should be constant and temperature-independent. The derived value of $S_H$ $\sim$ 0.22(2) V$^{-1}$ [Fig. 6(c)] is comparable with those in traditional itinerant ferromagnets, such as Fe and Ni ($S_H \sim 0.01 - 0.2$ V$^{-1}$).\cite{Dheer, Jan} It should be noted that the experimental data deviate from a linear fit below 30 K, which is probably related to magnetic disorder scattering induced resistivity upturn at low temperatures. Furthermore, the scaling plots of the modified anomalous Hall resistivity $\rho_{xy}/\mu_0H_{eff}$ and longitudinal resistivity $\rho_{xx}^2M/\mu_0H_{eff}$ over the whole temperature-magnetic-field range are shown in Fig. 6(d). To clarify, the curves in Fig. 6(d) have been offset subsequently by 0.02 cm$^3$ C$^{-1}$. The good linear behavior at different temperatures further confirm the conclusion that the AHE in Fe$_{3-x}$GeTe$_2$ is well described by the intrinsic KL theory.

When compared to CVT-grown crystals with $T_c$ of 220 K, we note that interatomic distances of the first coordination sphere Fe1-Fe2, Fe1-Ge, and Fe1-Te are smaller in our flux-grown crystals with $T_c$ of 153 K.\cite{Deiseroth} This is consistent with removal of Fe2 atoms and points to conclusion that significant weakening of FM interactions does not change dominant mechanism of AHE despite promotion of bad metal behavior.\cite{WangY}

\section{CONCLUSIONS}

In summary, we investigated the AHE in flux-grown Fe$_{3-x}$GeTe$_2$ ($x \approx 0.36$) crystals where significant amount of defect produces bad metallic behavior. The linear relationship between $\rho_{xy}/\mu_0H_{eff}$ and $\rho_{xx}^2M/\mu_0H_{eff}$ gives that the AHE in Fe$_{3-x}$GeTe$_2$ is dominated by the intrinsic KL mechanism. With the rapid development of 2D materials for spintronics, further investigation of AHE in the nano-sheet of Fe$_{3-x}$GeTe$_2$ is of high interest. It also motivates further studies to clarify the tolerability of Fe vacancies on designing spintronic devices.

\section*{Acknowledgements}
We thank John Warren for help with the SEM measurements. This work was supported by the U.S. Department of Energy (DOE)-BES, Division of Materials Science and Engineering, under Contract No. DE-SC0012704 (BNL). This research used the 8-ID (ISS) beamline of the National Synchrotron Light Source II, a U.S. DOE Office of Science User Facility operated for the DOE Office of Science by BNL under Contract No. DE-SC0012704.

\end{document}